\documentstyle[12pt]{article}
\newcommand{\h}{\hspace{.5cm}}
\newenvironment{destaque}{\begin{quotation}\small\em}{\end{quotation}}
\date{}
\title{Boundary Conditions as Mass Generation Mechanism for Real Scalar 
Fields}
\author{{\bf Jos\'e Alexandre Nogueira}\\
{\bf Pedro Leite Barbieri}\\
{\it Departamento de F\'{\i}sica, Centro de Ci\^encias Exatas,}\\
{\it Universidade Federal do Espirito Santo,}\\
{\it 29.060-900 - Vit\'oria-ES - Brazil,}\\
{\it E-mail: } nogueira@cce.ufes.br}
\begin{document}
\maketitle
\begin{abstract}
\begin{destaque}
We consider the effects of homogeneous Dirichlet's boundary conditions
 on two infinite parallel plane surfaces separated by some
 small distance
 {\it a}. We find that although spontaneous symmetry breaking does not
 occur for the theory of a massless, quartically self-interacting real
 scalar  field, the theory becomes a theory of a massive scalar field.

\end{destaque}
\end{abstract}
\section{Introduction}
\h The origin of particle mass has been a puzzle for theoretical
 physicists. In the contest of the standard model, the Higgs mechanism
 is well accepted as the mass generation mechanism. The underlying idea
 is that the universe is filled with a field, the Higgs field. The
 spin-zero Higgs field is a doublet in the SU(2) space and carries
 non-zero hypercharge, and it is a singlet in the SU(3) space of color.

Bosonic gauge and fermionic matter fields acquire mass from their 
interactions with the Higgs field. It is of fundamental importance to 
this mechanism that excited states (i. e., with one or more Higgs) are 
not orthogonal to the ground state (i. e., vacuum). Since states with 
one or more Higgs carry non-zero SU(2) and U(1) quantum numbers then
 they are non-zero for the vacuum as well. As a consequence of this 
the SU(2) and U(1) symmetries are spontaneously broken. The Lagrangian
 is symmetric under SU(2) and U(1) transformations but the vacuum is not \cite{ref1}.
 Technically this is achieved introducing a Higgs potential
 of imaginary mass and quartic interaction.

In the Coleman-Weinberg alternative approach the spontaneous symmetry 
breaking is induced by 1-loop radiative corrections and the mass being
 vanishing in the tree approximation \cite{ref2}. In the massless scalar 
electrodynamics, although the vacuum expectation value of the classical 
potential is unique (classically $<\phi> = 0$), in its renormalized form
 the effective potential in 1-loop becomes degenerate ($<\phi>$ is 
arbitrary). The mass renormalization condition fixes a determined 
vacuum and breaks the symmetry of the theory. The spontaneous symmetry 
breaking leads both scalar and vector particles to dynamically acquire
 mass whose value is proportional to the vacuum expectation value. 
This is only possible because there are two parameters in the theory:
 the electromagnetic coupling constant, $e$, and the coupling constant 
of the quartic self-interaction, $\lambda$, which must be of order
 $e^{4}$.

In a theory of massless real scalar field the existence of just one 
parameter does not allow a non-zero vacuum expectation value and the
 theory remains massless.

Boundary conditions do not affect the classical potential, but quantum 
corrections are affected and the effective potential changes. This may
 cause effects on the quantum vacuum expectation value $<\phi>$. 
Moreover boundary conditions introduce a new parameter, the length 
of the finite region. With these considerations we investigate the 
effects of boundary conditions on the vacuum expectation value 
$\left\langle \phi \right\rangle = 0$ and the mass generation. 
Besides the introduction of a new parameter makes more attractive 
the real scalar field theory.

The study of boundary conditions effects in Quantum Field Theory is
 not new. The Casimir effect \cite{ref3}, which is the change of the
 vacuum energy density due to constraints on the quantum field, 
induced by boundary conditions in space-time was experimentally 
observed in 1958 \cite{ref4} and recently \cite{ref5, ref6}, and 
a lot of applications have been accomplished [7 - 10],
 such as: gravity models, black holes, 
bag models, nonlinear meson theories describing baryons
 as solitons, the cosmological constant problem,
 compactification of the extra dimensions in Kaluza-Klein theories,
 quantum liquids \cite{ref11}, condensed matter \cite{ref12}.

In this work we study the effects of the boundary conditions on a 
massless real scalar field with quartic self-interaction which satisfies 
homogeneous Dirichlet conditions on two infinite parallel plane surfaces 
separated by some small distance $a$. The study of the $\lambda\phi^{4}$
 theory is very important in face of its applications in the 
Weinberg-Salam model of weak interactions, fermions masses generation, 
in solid state physics \cite{ref13}, inflationary models
 \cite{ref14}, solitons \cite{ref15} and Casimir effect \cite{ref16}.

It is important to stress that we do not use the imaginary mass term
 of the Higgs potential in order to avoid that it induces spontaneous 
symmetry breaking. Therefore with the purpose to evaluate the actual
 effects of boundary conditions we consider the real scalar field, 
with a second parameter introduced by the boundary conditions.

As will be seen subsequently the boundary conditions do not induce a 
degenerate vacuum state, although there is a typical length scale 
of the finite region where the massless scalar field acquires mass.

The outline of the paper is as follows: Section 2 is a brief review 
of how to calculate the effective potential at 1-loop. In section 3 we 
calculate the  effective potential for massless real scalar field, and 
evaluate the vacuum state and the renormalized mass. In section 4 we 
point out our conclusions and some  speculations.

\section{Effective Potential}

\h In the functional method approach of quantum field theory,  the
 effective potential is found as a loop expansion (or equivalently
 in powers of $\hbar$), that is, its classical amount plus quantum 
corrections [17 - 23].

Let $\phi(x)$ be a single real scalar field in a  Minkowski
space-time, subjected to the potential $V(\phi)$. The effective 
potential to the first order in the loop expansion is given by 

\begin{equation}
V_{ef}\left( \phi _{c}\right) =
 V_{cl}\left( \phi _{c}\right) +\frac{1}{2}\frac{\hbar }{\Omega }
\ln \det \left[ \frac{\delta ^{2}S\left[ \phi_{c}\right] }
{\delta \phi \left( x\right) \delta \phi \left( y\right) }
\right] = V_{cl}\left( \phi _{c}\right)
 + V_{ef}^{\left( 1\right) }\left( \phi_{c}\right) ,
\end{equation}
where the classic field $\phi _{c}$ is the vacuum expectation value
 in the presence of an external source $ J\left( x\right) $, taken 
as a constant value $\phi _{c}=\rho $, therefore, in the limit
 $ J\rightarrow 0 $, $ S\left[ \phi\right] $ is the classic 
action and $\Omega $ is the four dimensional space-time volume.

Performing the analytic continuation to the Euclidean space-time 
[18 - 20], the classical action can be written as

\begin{eqnarray}
S[\phi]= \int d^{4}x\biggl[\frac{1}{2}\partial_{\mu}\phi
\partial_{\mu}\phi
+ V_{cl}(\phi)\biggr],
\end{eqnarray}
where the Euclidean summation convention is assumed for repeated
indexes. From Eq.(2) we get the matrix ${\it m}(x,y)$ of the 
quadratic variation of the action $S[\phi]$

\begin{eqnarray}
{\it m}(x,y) \equiv \frac{\delta^{2}S[\phi]}{\delta\phi(x)
\delta\phi(y)} = \delta^{4}(x-y)[-\delta^{\mu\nu}\partial_{\mu}
\partial_{\nu} + V_{cl}^{\prime\prime}(\phi)].
\end{eqnarray}

Due to the Euclidean analytic continuation the operator {\it m} 
is a real, elliptical and self-adjoint. For operators with these 
properties we define a generalized zeta function. If $\{\lambda_{i}\}$
 are the eigenvalues of the operator $m(x,y)$, then the generalized
 zeta function associated to ${\it M}(x,y)$
 $(m \rightarrow {\it M} = \frac{m}{\mu^{2}})$ is defined by 

\begin{eqnarray}
\zeta_{\it M}(s) = \sum_{i}\biggl(
\frac{\lambda_{i}}{\mu^{2}}\biggr)^{-s},
\end{eqnarray}
where we have introduced an unknown scale parameter $\mu$, 
with dimensions of (length)$^{-1}$ or mass in order to keep the
 zeta function dimensionless for all s. The introduction of the scale 
parameter $\mu$, can be best understood when we observe that a hidden 
splitting of the divergent integral there is in the proceeding of zeta 
function regularization, that is, a separation of the divergent 
and finite parts of the
$V_{ef}(\phi_{c})$ (in \cite{ref20}, page 208 and in \cite{ref21},
 page 88). It is well-known the relation \cite{ref18} 

\begin{eqnarray}
\ln\det{\it M} = - \frac{d\zeta_{\it M}(0)}{ds}.
\end{eqnarray}

Now, the effective potential in the one-loop approximation can
 be written as

\begin{eqnarray}
V_{ef}({\phi}_{c}) = V_{cl}(\phi_{c}) -
 \frac{1}{2}\frac{\hbar }{\Omega }\frac{d\zeta_{\it M}(0)}{ds}.
\end{eqnarray}

Due to the regularity of the generalized zeta function at $s=0$
 \cite{ref24}, the evaluation of the effective potential (Eq.(6)) 
gives a finite result with no need of substraction of any pole, 
or addition of infinite counter-terms. Evidently, the fitting of 
the theory parameters taking into account the observed results, 
leads to the renormalization conditions
\begin{equation}
\left. \frac{d^{2}V_{ef}}{d\phi _{c}^{2}}\right| _{\phi _{c}=
\left\langle \phi \right\rangle }=m_{R}^{2},
\end{equation}
\begin{equation}
\left. \frac{d^{4}V_{ef}}{d\phi _{c}^{4}}\right| _{\phi _{c}=
\left\langle \phi \right\rangle }=\lambda _{R},
\end{equation}
where $m_{R}$ is the renormalized mass, $\lambda_{R}$ is the 
renormalized coupling constant and $<\phi>$ is the minimum of 
the effective potential (subtraction or renormalization point)
 \cite{ref22}. Since $\phi_{c}$ takes on value $<\phi>$ in the 
ground state, then $<\phi>$ is called the vacuum expectation 
value of $\phi$, $<\phi> = <0|\phi|0>$.

In the case of theories of null mass, the subtraction point for the
renormalization condition (8) cannot be taken at $\left\langle \phi
\right\rangle =0$ due to the logarithmic singularity. Even so in that
 case there is no intrinsic mass scale; therefore all the 
renormalization points are equivalent and the condition (8) is 
replaced by 
\begin{equation}
\left. \frac{d^{4}V_{ef}}{d\phi _{c}^{4}}\right| _{\phi _{c}=M}
=\lambda _{R},
\end{equation}
where $M$ is a arbitrary floating renormalization point \cite{ref20}.

\section{Mass Generation}

\h Now, let us consider the theory of a massless, quartically 
self-interacting real scalar field $\phi(x)$ satisfying homogeneous 
Dirichlet's boundary conditions on two infinite parallel plane surfaces 
separated by some small distance $a$.

The Lagrangian density for this theory is
\begin{equation}
{\cal L=}\frac{1}{2}\partial _{\mu }\phi \partial ^{\mu }\phi
 -\frac{\lambda }{4!}\phi ^{4}.
\end{equation}
The Lagrangian above is even in $\phi$, so it is invariant under
 discreet symmetry (G-parity) defined by the transformation 
$\phi \rightarrow -\phi$.

In this case the zeta function, defined by Eq.(4), is given by 
\begin{equation}
\zeta _{m}\left( s\right) =
\sum_{N=1}^{\infty }\int_{-\infty }^{+\infty }
\frac{\Omega }{\left( 2\pi \right) ^{3}a}d^{3}k\left[
 k^{2}+\frac{\pi^{2}N^{2}}{a^{2}}+\frac{\lambda }{2}
\phi _{c}^{2}\right] ^{-s}.
\end{equation}
Using the integral \cite{ref25} 
\begin{equation}
\int_{-\infty }^{+\infty }\left[ k^{2}+A^{2}\right] ^{-s}d^{m}k=
\frac{\pi ^{\frac{m}{2}}\Gamma \left( s-\frac{m}{2}\right) }
{\Gamma \left( s\right) }\left( A^{2}\right) ^{\frac{m}{2}-s},
\end{equation}
and using the formula \cite{ref26} 
\begin{equation}
\sum_{N=1}^{\infty }\left[ N^{2}+B^{2}\right] ^{-p}=
-\frac{1}{2}B^{-2p}+
\frac{\pi ^{\frac{1}{2}}}{2B^{2p-1}\Gamma \left( p\right) }
\left[ \Gamma \left( p-\frac{1}{2}\right)
 +4\sum_{N=1}^{\infty }\frac{K_{p-\frac{1}{2}}
\left( 2\pi NB\right) }{\left( N\pi B
\right)^{\frac{1}{2}-p}}\right] ,
\end{equation}
where $K_{\nu }(x)$ are modified Bessel functions, we get 
$$
\zeta _{m}\left( s\right) =-\frac{\Omega }{16\pi ^{\frac{3}{2}}a}
\frac{\Gamma \left( s-\frac{3}{2}\right) }{\Gamma \left( s\right) }
\left(\alpha\phi _{c}\right)^{3-2s} +
$$
$$
+ \frac{\Omega }{16\pi ^{2}}\frac{1}{\left(s-1\right)
 \left( s-2\right) }\left( \alpha\phi _{c}\right)^{4-2s} +
$$
\begin{equation}
+ \frac{\Omega }{4\pi ^{2}}\frac{\left( \alpha
\phi_{c}\right)^{4-2s}}{\Gamma \left( s\right) }
\sum_{N=1}^{\infty }\frac{K_{s-2}\left( 
2N\alpha a\phi _{c}\right) }
{\left( N\alpha a\phi _{c}\right) ^{2-s}},
\end{equation}
where we define $\alpha^{2}=\frac{\lambda }{2}$ for the sake 
of simplicity.

In order to calculate the one-loop effective potential, we compute
 $\zeta _{m}\left( 0\right)$ and $\zeta _{m}^{\prime }\left(0\right)$
 from Eq.(14) and use them in Eq.(6)$\footnote{%
We keep $
\rlap{\protect\rule[1.1ex]{.325em}{.1ex}}h%
$ to mark the quantum corrections, but we set
 $
\rlap{\protect\rule[1.1ex]{.325em}{.1ex}}h%
=c=1$ everywhere else.}$
$$
V_{ef}\left( \phi _{c}\right) = \frac{2\alpha^{2}}{4!}\phi_{c}^{4} + 
\frac{\hbar}{24\pi a}\alpha ^{3}\phi_{c}^{3} +
\frac{\hbar }{64\pi^{2}}\alpha^{4}\phi _{c}^{4}
\left[ \ln\left(\frac{\alpha^{2} \phi _{c}^{2}}{\mu ^{2}}\right) - 
\frac{3}{2} \right] +
$$
\begin{equation}
- \frac{\hbar}{8\pi ^{2}}\alpha ^{4}\phi _{c}^{4}\sum_{N=1}^{\infty }
\frac{K_{2}\left( 2N\alpha a\phi _{c}\right)}
{\left( N\alpha a\phi_{c}\right) ^{2}},
\end{equation}
recalling that $K_{\nu }(x)=K_{-\nu }(x)$. We  notice that, as we have
 to take absolute value of $\phi_{c}$ in the formula (13), the second 
term on the right-hand side of Eq.(15) does not yield symmetry breaking.

Let $m \geq 0$ be an integer number, such that,
 $2m\alpha a\phi_{c} < 1$ and $2(m+1)\alpha a\phi_{c} \geq 1$.
 Then, the sum in Eq.(15) can be write as
$$
\frac{\hbar }{8\pi ^{2}}\alpha ^{4}\phi_{c}^{4}
\sum_{N=1}^{\infty}\frac{K_{2}\left( 2N\alpha a\phi _{c}\right) }
{\left( N\alpha a\phi _{c}\right) ^{2}} =
$$
\begin{equation}
= \frac{\hbar }{8\pi ^{2}}\alpha ^{4}\phi_{c}^{4}
\sum_{N=1}^{m}\frac{K_{2}\left( 2N\alpha a\phi _{c}\right) }
{\left( N\alpha a\phi _{c}\right) ^{2}} 
+ \frac{\hbar }{8\pi ^{2}}\alpha ^{4}\phi_{c}^{4}
\sum_{N=m+1}^{\infty}\frac{K_{2}\left( 2N\alpha a\phi _{c}
\right) } {\left( N\alpha a\phi _{c}\right) ^{2}}.
\end{equation}
Since $2N\alpha a\phi_{c} < 1$, for any $N \leq m$, we expand the 
first term on the right-hand side of Eq.(16) using the relation
 \cite{ref27, ref31}
$$
\left( \frac{x}{2}\right)^{\nu}K_{\nu }\left( x\right) 
=\frac{1}{2}\sum_{a=0}^{\nu -1}\left( -1\right) ^{a}\left( \frac{x}{2}
\right)^{2a}\frac{\Gamma \left( \nu -a\right)}
{\Gamma \left( a+1\right)} +
$$
\begin{equation}
 + \sum_{a=0}^{\infty }\frac{\left( -1\right) ^{\nu }\left( 
\frac{x}{2}\right)^{2\nu +2a}}{\Gamma \left( a+1\right) \Gamma
 \left( \nu +a+1\right) }\left[ \psi \left( a+1\right)
 +\psi \left( \nu +a+1\right) -2\ln \left( \frac{x}{2}\right) \right] ,
\end{equation}
for $\nu > 0$, to get
$$
\frac{\hbar }{8\pi ^{2}}\alpha ^{4}\phi_{c}^{4}
\sum_{N=1}^{\infty}\frac{K_{2}\left( 2N \alpha a\phi_{c}\right)}
{\left( N\alpha a\phi _{c}\right) ^{2}}=
\frac{\hbar}{16\pi^{2}a^{4}}\sum_{N=1}^{m}\left(\frac{1}{N}
\right)^{4} - 
\frac{\hbar}{16\pi^{2}a^{2}}\alpha^{2}\phi_{c}^{2}\sum_{N=1}^{m}
\left(\frac{1}{N}\right)^{2} +
$$
$$
-\frac{\hbar}{16\pi^{2}}\alpha^{4}\phi_{c}^{4}\sum_{N=1}^{m}
\left[\ln\left(N\alpha a\phi_{c}\right)
 + \gamma - 3/4 \right] + \frac{\hbar}{8\pi^{2}} 
\alpha^{4}\phi_{c}^{4}\sum_{N=1}^{m}{\cal O}\left(N\alpha 
a\phi_{c}\right)^{2} +
$$
\begin{equation}
+ \frac{\hbar}{8\pi^{2}}
\alpha^{4}\phi_{c}^{4}\sum_{N=m+1}^{\infty}\frac{K_{2}\left(
 2N \alpha a\phi_{c}\right)}{\left( N\alpha a\phi _{c}\right) ^{2}},
\end{equation}
where $\gamma$ is the Euler number.

The first sum on the right-hand side of Eq.(16) exists, provided 
${\bar a}$ (${\bar a} = a \mu$) be of order $\alpha^{n}$ ($n \geq 0$),
 since ${\bar \phi_{c}}$ (${\bar \phi_{c}} = \frac{\phi_{c}}{\mu}$)
 must be of order $\alpha^{0}$. Hence, up to higher-order terms, the effective potential becomes
\begin{equation}
V_{ef}\left( \phi _{c}\right) = \frac{2\alpha^{2}}{4!}\phi_{c}^{4} +
\frac{\hbar}{24\pi a}\alpha ^{3}\phi_{c}^{3} -
\frac{\hbar}{16\pi^{2}a^{4}}\sum_{N=1}^{m}\left(
\frac{1}{N}\right)^{4} + \frac{\hbar}{16\pi^{2}a^{2}}\alpha^{2}\phi_{c}^{2}\sum_{N=1}^{m}
\left(\frac{1}{N}\right)^{2}.
\end{equation}
For large $m$, we may approximate the two sums in Eq.(19) to 
Riemann zeta functions and to obtain
\begin{equation}
V_{ef}\left( \phi _{c}\right) =\frac{2\alpha^{2}}{4!}\phi_{c}^{4}
+ \frac{\hbar}{24\pi a}\alpha ^{3}\phi_{c}^{3} - \frac{\hbar\pi^{2}}{1440a^{4}}
+ \frac{\hbar\alpha ^{2}\phi _{c}^{2}}{96a^{2}}.
\end{equation}

The minimum occurs at $\phi _{c}=\left\langle \phi \right\rangle $, 
where 
\begin{equation}
\left. \frac{dV_{ef}}{d\phi _{c}}\right| _{\phi _{c}=
\left\langle \phi \right\rangle }=0.
\end{equation}
Differentiating Eq.(20), we have
\begin{equation}
\left\langle \phi \right\rangle \left[\frac{\alpha^{2}}{3}
\left\langle \phi \right\rangle^{2} + \frac{\hbar\alpha^{3}}
{8\pi a}\left\langle \phi \right\rangle + \frac{\hbar\alpha ^{2}}
{48a^{2}} \right] = 0.
\end{equation}
There will be non-trivial solution of Eq.(22) if the sum of the terms
 between brackets vanishes. However there is not $\bar{a}$ (with respect
 to the orders of $\alpha$) which satisfies Eq.(22). Therefore the
 minimum of $V_{ef}$ is satisfied for
\begin{equation}
\left\langle \phi \right\rangle = 0.
\end{equation}

Eq.(23) shows that the vacuum is non-degenerate, and therefore 
spontaneous symmetry breaking cannot occur. Actually, the boundary
 conditions allow just one constant vacuum solution,
  $\left\langle \phi \right\rangle = 0$. If the result had been 
$\left\langle \phi \right\rangle = constant \neq 0$,
 the effective potential would not have been used and a solution
$\left\langle \phi \right\rangle = \phi_{0}(x)$ would have been
 expected because the boundary conditions break the translational
 invariance.

Although Eq.(20) is finite, that is not a final result because the
 coupling constant in it is an arbitrary parameter. Therefore we must
 fit it to the renormalized coupling constant using the renormalization 
condition Eq.(8) to get $\lambda = \lambda_{R}$.

The renormalized mass of the scalar field is given by
\begin{equation}
\left. \frac{d^{2}V_{ef}}{d\phi _{cl}^{2}}\right| 
_{\phi_{c}=\left\langle\phi\right\rangle = 0 }=\frac{\hbar 
\lambda_{R}}{96a^{2}}=m_{R}^{2},
\end{equation}
that unlike what we expected it can be non-zero.

Eq.(24) shows if ${\bar a}^{2}$ is of order $\lambda$ then the mass 
will be of order $\lambda^{0}$. This result is in agreement with our 
initial assumption. If ${\bar a}^{2}$ is of lower order than 
$\lambda$ then the mass will lie far outside the expected range of
 validity of our approximation. It follows that there is a typical
 length scale given by the parameter $a$ of the theory. Within this 
length scale, the massless scalar field theory with self-interaction
 becomes massive due to boundary conditions effects.

Although topological mass has already been obtained \cite{ref32, ref34} to order $\lambda$ for a theory which is massless at the tree-graph level, 
we stress the fact that if $a$ is small enough the theory will become 
one massive in order $(\hbar^{0})$. This is because $\lambda$ and $a$ are 
independent parameters, so the mass term found can be of order zero-loop,
even though it is a one-loop result (it is from radiative corrections).
Our result of Eq.(24) is in agreement with that obtained by David J. Tom \cite{ref34}.

\section{Conclusion}

\h We have studied the theory of a massless, quartically 
self-interacting real scalar field satisfying homogeneous 
Dirichlet's boundary conditions on two infinite parallel 
plane surfaces separated by some small distance $a$. As a  result, 
we infer:\\
i) For small $a$ (${\bar a} \propto \lambda^{n}$, $n > 0$) there is 
not any order of ${\bar a}$ with regard to order of $\lambda$ which 
leaves the ground state (vacuum) degenerate, i. e., spontaneous 
symmetry breaking does not occur. So, the effective potential can be used to evaluate the renormalized mass and the Casimir energy.\\
ii) There is a typical length scale of the finite region where 
massless scalar field acquires mass, i. e.,
 if ${\bar a} \propto\lambda^{n}$, with $n \geq 1$, the 
theory becomes a theory of a massive 
real scalar field. Therefore, when $a$ becomes small enough the theory 
undergoes a transition from one massless to one massive.\\
iii) Since spontaneous symmetry breaking does not occur, the mass 
generation is only due to the boundary condition.\\

Finally,  we speculate that boundary conditions may be a mechanism
 of mass generation, i. e., that massless theories defined in finite
 regions of space-time become massive theories. Therefore we conjecture 
that confinement may be a candidate for an alternative mechanism for
 the mass generation of quarks.

It is clear we do not claim that these are actual boundary conditions 
which produce the masses of the particles. We only consider boundary 
condition may be an alternative mechanism for particle mass generation.\\

{\bf Acknowledgments}

\h J. A. Nogueira is especially grateful to Professor Olivier Piguet
 who has helped to clarify many obscure points. We would like to 
express our thanks to Dr. Manoelito Martins de Souza  and 
to Dr. Francisco de Assis Ribas Bosco. It is a pleasure to thank 
Dr. Luiz C. Albuquerque for helpful comments. This work was supported 
in part by the National Agency for Research (CAPES)(Brazil).\\

\end{document}